\documentclass[12pt,a4]{article}
\usepackage{amsmath}
\usepackage{bm}
\usepackage{amsfonts}
\usepackage{amssymb}
\usepackage{graphics}
\usepackage[normal]{caption2}
\usepackage{subfigure}
\usepackage{rotating}
\usepackage{citesort}
\def\be{\begin{equation}}
\def\ee{\end{equation}}
\def\ba{\begin{array}}
\def\ea{\end{array}}
\def\bea{\begin{eqnarray}}
\def\eea{\end{eqnarray}}

\begin{document}
\baselineskip 20pt \setlength\tabcolsep{2.5mm}
\renewcommand\arraystretch{1.5}
\setlength{\abovecaptionskip}{0.1cm}
\setlength{\belowcaptionskip}{0.5cm}
\pagestyle{empty}
\newpage
\pagestyle{plain} \setcounter{page}{1} \setcounter{lofdepth}{2}
\begin{center} {\large\bf Role of colliding geometry on the N/Z dependence of balance energy}\\
\vspace*{0.4cm}

{\bf Sakshi Gautam$^a$}, {\bf Aman D. Sood$^b$} and {\bf Rajeev K. Puri$^{a}$}\footnote{Email:~amandsood@gmail.com}\\
$^a${\it  Department of Physics, Panjab University, Chandigarh
-160 014, India.\\}$^{b}$ {\it  SUBATECH, Laboratoire de Physique
Subatomique et des Technologies Associ\'{e}es, Universit\'{e} de
Nantes - IN2P3/CNRS - EMN 4 rue Alfred Kastler, F-44072 Nantes,
France.\\}
\end{center}
We study the role of colliding geometry on the N/Z dependence of
balance energy using isospin-dependent quantum molecular dynamics
model. Our study reveals that the N/Z dependence of balance energy
becomes much steeper for peripheral collisions as compared to the
central collisions. We also study the effect of system mass on the
impact parameter dependence of N/Z dependence of balance energy.
The study shows that lighter systems shows greater sensitivity to
colliding geometry towards the N/Z dependence.
\newpage
\baselineskip 20pt
\section{Introduction}

The construction of radioactive ion beams (RIBs) facilities around
the world has generated a lot of interest in isospin physics
\cite{rib1,rib2,rib3}. These facilities provide the opportunities
to study the nuclear reactions involving nuclei with neutron or
proton excess. These studies are helpful in investigating the
structure of rare isotopes and the properties of isospin
asymmetric nuclear matter. The ultimate goal of isospin physics by
heavy-ion collisions of neutron-rich radioactive nuclei is to
explore the isospin dependence of in-medium nuclear effective
interactions and the equation of state of asymmetric nuclear
matter.
 \par
During the last few decades there have been significant activities
in exploring the isospin effects in collective flow
\cite{li96,pak97,gaum1} and multifragmentation
\cite{fang00,wang09}. Various studies have been done in recent
past to investigate the isospin effects in collective flow and in
its disappearance (at a particular incident energy called balance
energy E$_{bal}$) \cite{li96,pak97,gaum1,gaum2,gaum3}. The isospin
effects in collective flow have been explained in literature as
the competition
 among various reaction mechanisms, such as nucleon-nucleon collisions, symmetry energy, surface properties and Coulomb
 force. The relative importance among these reaction mechanisms
 is not yet clear \cite{li96}. To shed light on the relative importance of above mentioned
 mechanisms, in Ref.
\cite{gaum2,gaum3} Gautam \emph{et al}. have studied the isospin
effects in E$_{bal}$ and its system size dependence throughout the
range of colliding geometry. The study pointed towards the
dominance of Coulomb potential in isospin effects at all the
colliding geometries. Moreover, the study also pointed that the
effect of symmetry energy is uniform throughout the mass range and
range of colliding geometries. So to look for an observable which
could show the sensitivity to symmetry energy, Sood has studied
the N/Z dependence of E$_{bal}$ for isotopic series of Ca
\cite{gaum4}. The study shows that the N/Z dependence of E$_{bal}$
is sensitive to symmetry energy and shows insensitivity towards
the isospin dependence of nucleon-nucleon cross section and
henceforth the study revealed that the N/Z dependence of symmetry
energy can act as a probe of symmetry energy. To see the system
size effects on N/Z dependence of E$_{bal}$, followed by this,
Gautam and Sood carried out the above mentioned study throughout
the mass range at semicentral colliding geometry of b/b$_{max}$ =
0.2-0.4 \cite{gaum5}. Their study showed that the sensitivity of
N/Z dependence of E$_{bal}$ is more for lighter systems. In the
present paper, we plan to extend the above study to the whole
range of colliding geometries. For the present study we use
isospin-dependent quantum molecular dynamics (IQMD) model
\cite{hart98,aichqmd}.
\section{The Model}
 The IQMD model is an extension of the QMD model \cite{aichqmd}, which treats different charge states of
nucleons, deltas and pions explicitly, as inherited from the
Vlasov-Uehling-Uhlenbeck (VUU) model \cite{vuu}. The IQMD model
has been used successfully for the analysis of a large number of
observables from low to relativistic energies. The isospin degree
of freedom enters into the calculations via symmetry potential,
cross sections and Coulomb interaction.
 \par
 In this model, baryons are represented by Gaussian-shaped density distributions
  \begin {eqnarray}
  f_{i}(\vec{r},\vec{p},t) =
  \frac{1}{\pi^{2}\hbar^{2}}\exp(-[\vec{r}-\vec{r_{i}}(t)]^{2}\frac{1}{2L})
   \times \exp(-[\vec{p}- \vec{p_{i}}(t)]^{2}\frac{2L}{\hbar^{2}})
 \end {eqnarray}
 Nucleons are initialized in a sphere with radius R = 1.12 A$^{1/3}$ fm, in accordance with the liquid-drop model.
 Each nucleon occupies a volume of \emph{h$^{3}$}, so that phase space is uniformly filled.
 The initial momenta are randomly chosen between 0 and Fermi momentum ($\vec{p}$$_{F}$).
 The nucleons of the target and projectile interact by two- and three-body Skyrme forces, Yukawa potential and
  Coulomb interactions. In addition to the use of explicit charge states of all baryons and mesons, a symmetry potential between protons and neutrons
 corresponding to the Bethe-Weizsacker mass formula has been included.
 The hadrons propagate using the Hamilton equations of motion:
 \begin {eqnarray}
 \frac{d\vec{{r_{i}}}}{dt} = \frac{d\langle H
 \rangle}{d\vec{p_{i}}};\frac{d\vec{p_{i}}}{dt} = - \frac{d\langle H
  \rangle}{d\vec{r_{i}}}
 \end {eqnarray}
  with
  \begin {eqnarray}
  \nonumber\langle H\rangle =\langle T\rangle
+ \langle V \rangle \\
                  \nonumber= \sum_{i}\frac{p^{2}_{i}}{2m_{i}} +
                   \sum_{i}\sum_{j>i}\int
                   f_{i}(\vec{r},\vec{p},t)V^{ij}
                   (\vec{r}~',\vec{r}) \\\times
                   f_{j}(\vec{r}~',\vec{p}~',t)
                   d\vec{r}~ d\vec{r}~'~ d\vec{p}~ d\vec{p}~'.
 \end {eqnarray}
 The baryon potential\emph{ V$^{ij}$}, in the above relation, reads as
 \begin {eqnarray}
  \nonumber V^{ij}(\vec{r}~'-\vec{r}) = V^{ij}_{Skyrme} + V^{ij}_{Yukawa} +
  V^{ij}_{Coul} + V^{ij}_{sym} \\
    \nonumber=[t_{1}\delta(\vec{r}~'-\vec{r})+t_{2}\delta(\vec{r}~'-\vec{r})\rho^{\gamma-1}(\frac{\vec{r}~'+\vec{r}}{2})]\\
   \nonumber ~+t_{3}\frac{\exp(|(\vec{r}~'-\vec{r})|/\mu)}{(|(\vec{r}~'-\vec{r})|/\mu)}+
    \frac{Z_{i}Z_{j}e^{2}}{|(\vec{r}~'-\vec{r})|}\\
       +t_{4}\frac{1}{\varrho_{0}}T_{3i}T_{3j}\delta(\vec{r_{i}}~'-\vec{r_{j}}).
 \end {eqnarray}
Here t$_{6}$ = 4C with C = 32 MeV and \emph{Z$_{i}$} and
\emph{Z$_{j}$} denote the charges of the \emph{ith} and \emph{jth}
baryon, and \emph{T$_{3i}$} and
 \emph{T$_{3j}$}
 are their respective \emph{T$_{3}$} components (i.e. $1/2$ for protons and $-1/2$ for neutrons).
The parameters\emph{ $\mu$} and \emph{t$_{1}$,....,t$_{4}$} are
adjusted to the real part of the nucleonic optical potential.
 For the density dependence of  the nucleon optical potential, standard Skyrme-type parametrization is employed.

\par

\section{Results and discussions}

\begin{figure}[!t]
\centering
 \vskip 1cm
\includegraphics[angle=0,width=12cm]{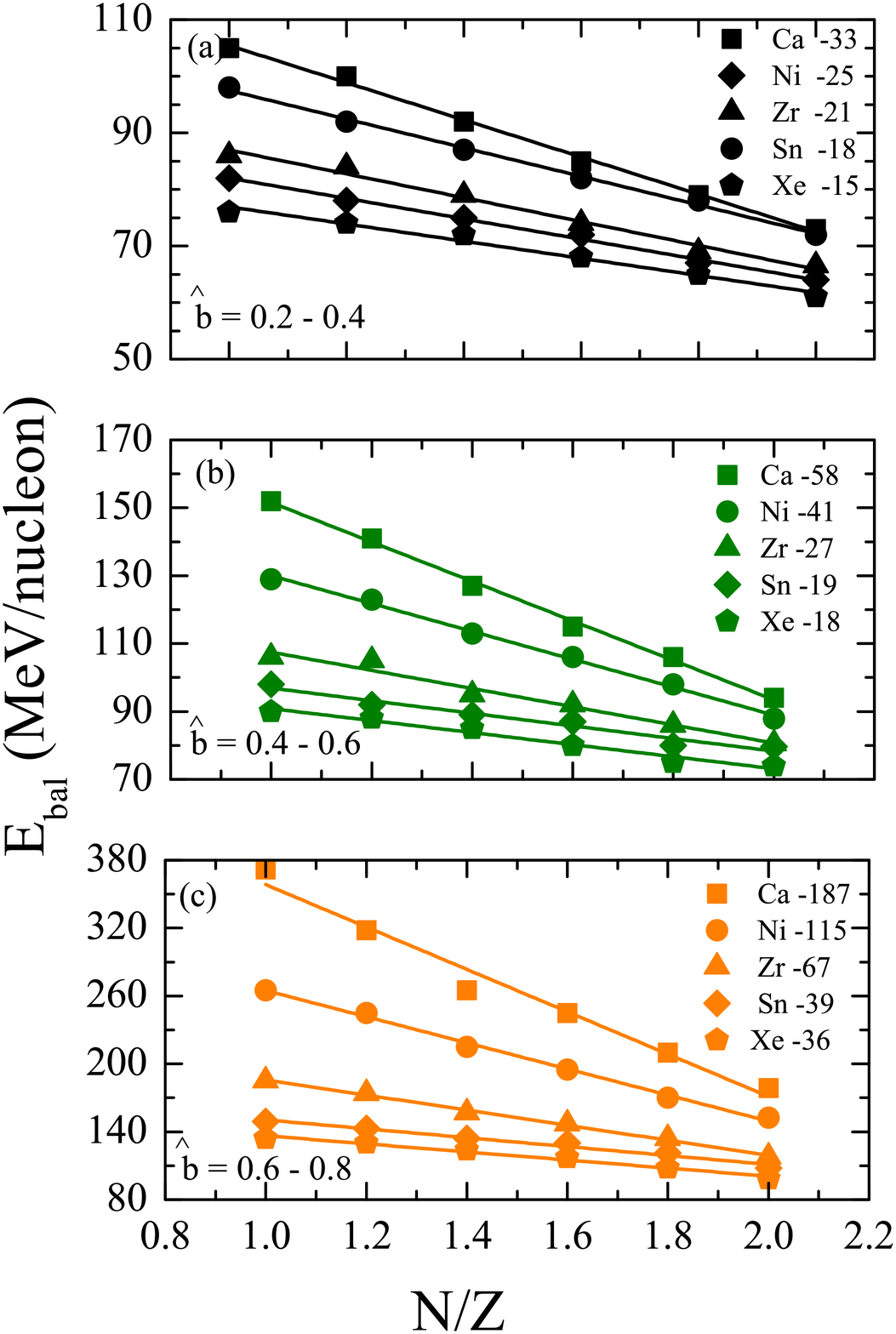}
 \vskip -0cm \caption{ N/Z dependence of E$_{bal}$ for various systems. Various symbols are explained
 in the text. Lines are linear fit. Top, middle and bottom panels represent the results for b/b$_{max}$ = 0.2 - 0.4, 0.4 - 0.6 and
0.6 - 0.8, respectively. }\label{fig1}
\end{figure}
\begin{figure}[!t]
\centering
 \vskip 1cm
\includegraphics[angle=0,width=12cm]{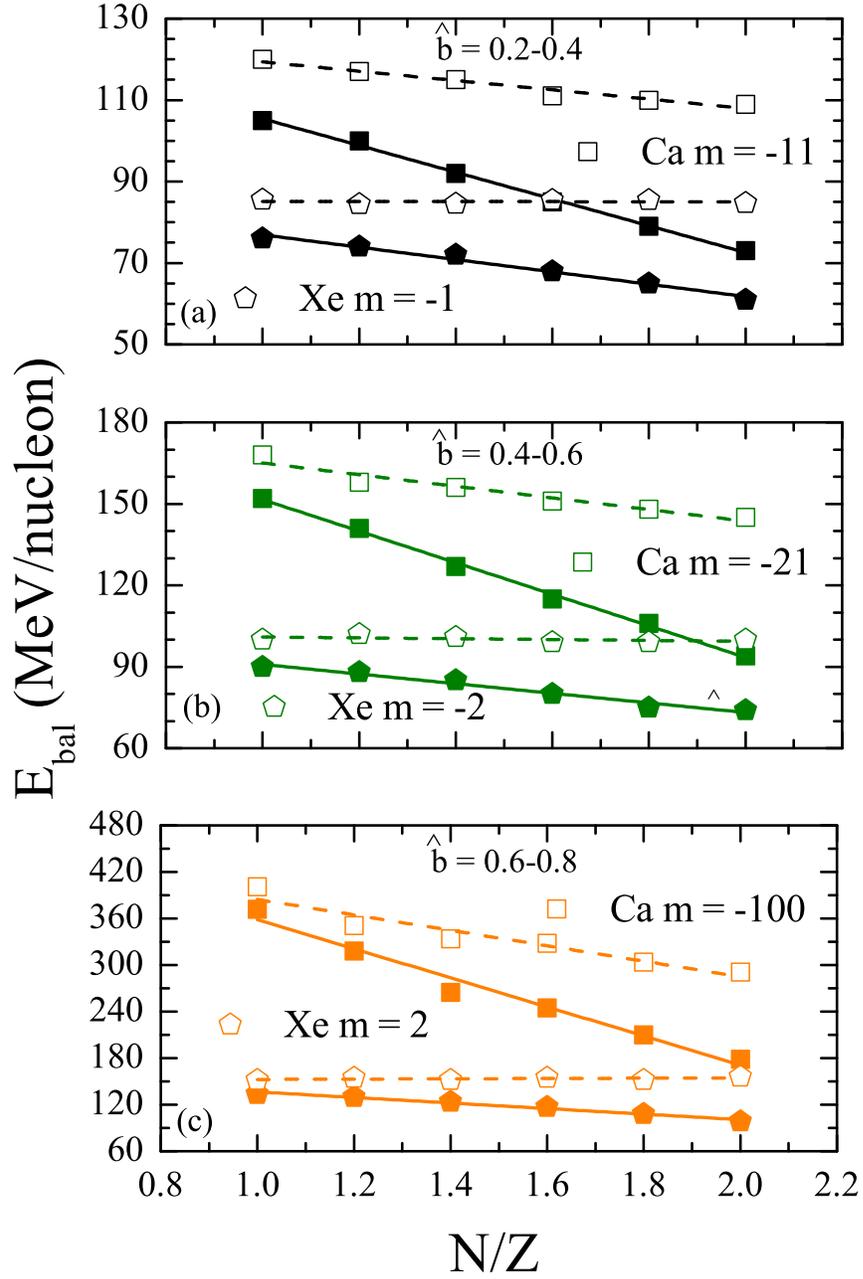}
 \vskip -0cm \caption{ N/Z dependence of E$_{bal}$ for Ca+Ca and Xe+Xe with E$_{sym}$ on and off. Various symbols are explained in text. Top, middle and bottom panels represent the results for b/b$_{max}$ = 0.2 - 0.4, 0.4 - 0.6 and
0.6 - 0.8, respectively.}\label{fig2}
\end{figure}

We simulate the reactions of Ca+Ca, Ni+Ni, Zr+Zr, Sn+Sn, and Xe+Xe
with N/Z varying from 1.0 to 2.0 in small steps of 0.2. In
particular we simulate the reactions of $^{40}$Ca+$^{40}$Ca,
$^{44}$Ca+$^{44}$Ca, $^{48}$Ca+$^{48}$Ca, $^{52}$Ca+$^{52}$Ca,
$^{56}$Ca+$^{56}$Ca, and $^{60}$Ca+$^{60}$Ca; $^{56}$Ni+$^{56}$Ni,
$^{62}$Ni+$^{62}$Ni, $^{68}$Ni+$^{68}$Ni, $^{72}$Ni+$^{72}$Ni, and
$^{78}$Ni+$^{78}$Ni; $^{81}$Zr+$^{81}$Zr, $^{88}$Zr+$^{88}$Zr,
$^{96}$Zr+$^{96}$Zr, $^{104}$Zr+$^{104}$Zr, and
$^{110}$Zr+$^{110}$Zr; $^{100}$Sn+$^{100}$Sn,
$^{112}$Sn+$^{112}$Sn, $^{120}$Sn+$^{120}$Sn,
$^{129}$Sn+$^{129}$Sn, and $^{140}$Sn+$^{140}$Sn; and
$^{110}$Xe+$^{110}$Xe, $^{120}$Xe+$^{120}$Xe,
$^{129}$Xe+$^{129}$Xe, $^{140}$Xe+$^{140}$Xe, and
$^{151}$Xe+$^{151}$Xe at b/b$_{max}$ = 0.2 - 0.4, 0.4 - 0.6 and
0.6 - 0.8. We also use a soft equation of state along with the
standard isospin- and energy-dependent cross section reduced by
  20$\%$, i.e. $\sigma$ = 0.8 $\sigma_{nn}^{free}$.
The details about the elastic and inelastic cross sections for
proton-proton and proton-neutron collisions can be found in
\cite{hart98,cug}. The cross sections for neutron-neutron
collisions are assumed to be equal to the proton-proton cross
sections. The reactions are followed till the transverse in-plane
saturates. In the present study we use the quantity
"\textit{directed transverse momentum $\langle
p_{x}^{dir}\rangle$}" which is defined as \cite{sood1,sood2}
\begin {equation}
\langle{p_{x}^{dir}}\rangle = \frac{1} {A}\sum_{i=1}^{A}{sign\{
{y(i)}\} p_{x}(i)},
\end {equation}
where $y(i)$ and $p_{x}$(i) are, respectively, the rapidity and
the momentum of the $i^{th}$ particle. The rapidity is defined as
\begin {equation}
Y(i)= \frac{1}{2}\ln\frac{{\vec{E}}(i)+{\vec{p}}_{z}(i)}
{{\vec{E}}(i)-{\vec{p}}_{z}(i)},
\end {equation}
where $\vec{E}(i)$ and $\vec{p_{z}}(i)$ are, respectively, the
energy and longitudinal momentum of the $i^{th}$ particle. In this
definition, all the rapidity bins are taken into account. It is
worth mentioning that the E$_{bal}$ has the same value for all
fragments types \cite{pak97,west93,west98,cuss}. Further the
apparatus corrections and acceptance do not play any role in
calculation of the E$_{bal}$ \cite{west93,cuss,ogli89}.

\begin{figure}[!t]
\centering
 \vskip 1cm
\includegraphics[angle=0,width=12cm]{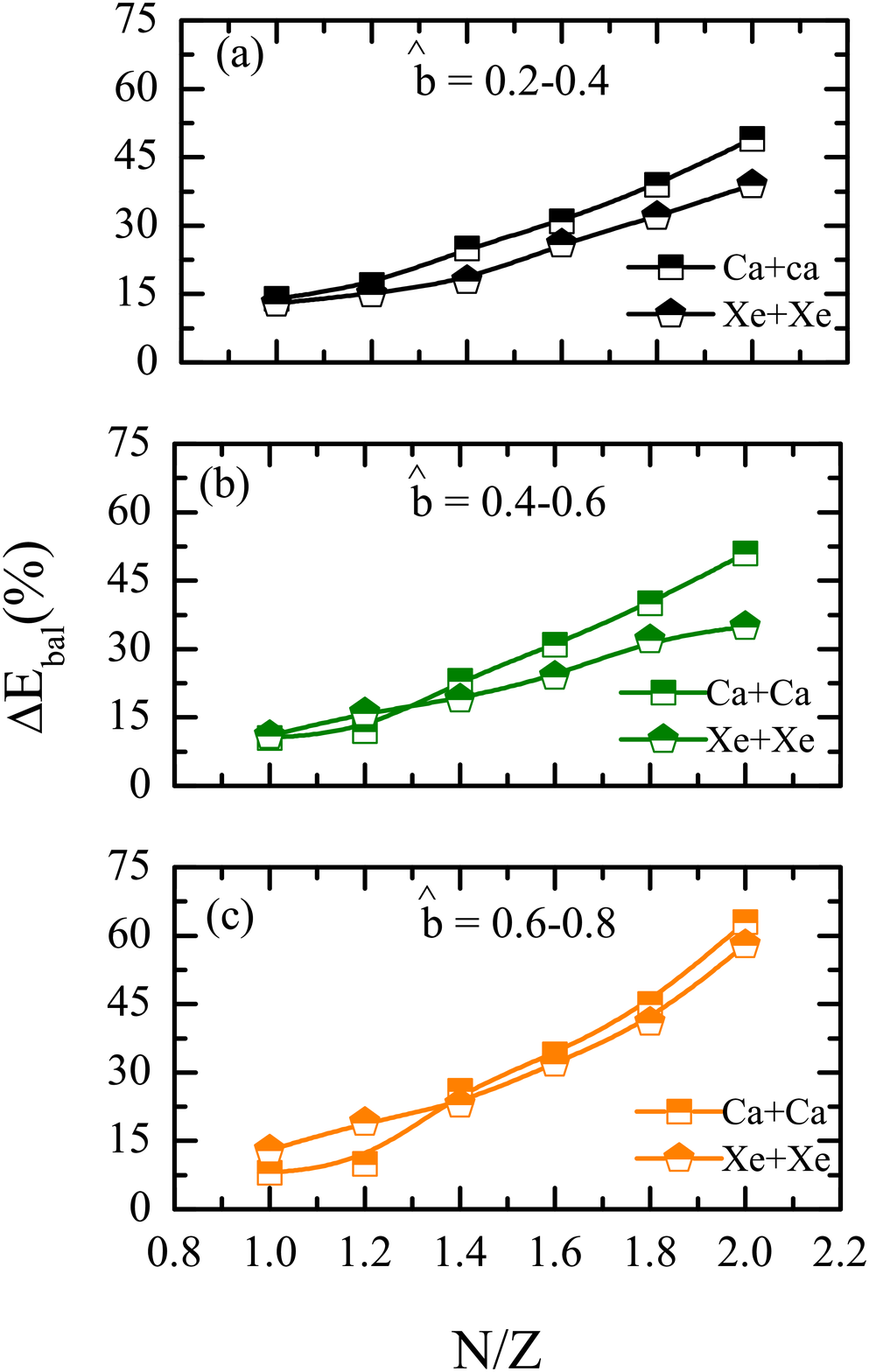}
 \vskip -0cm \caption{N/Z dependence of $\Delta$E$_{bal}$($\%$) for Ca and Xe series. Various symbols are explained in text. Lines are guide to the eye.}\label{fig3}
\end{figure}

\begin{figure}[!t]
\centering
 \vskip 1cm
\includegraphics[angle=0,width=14cm]{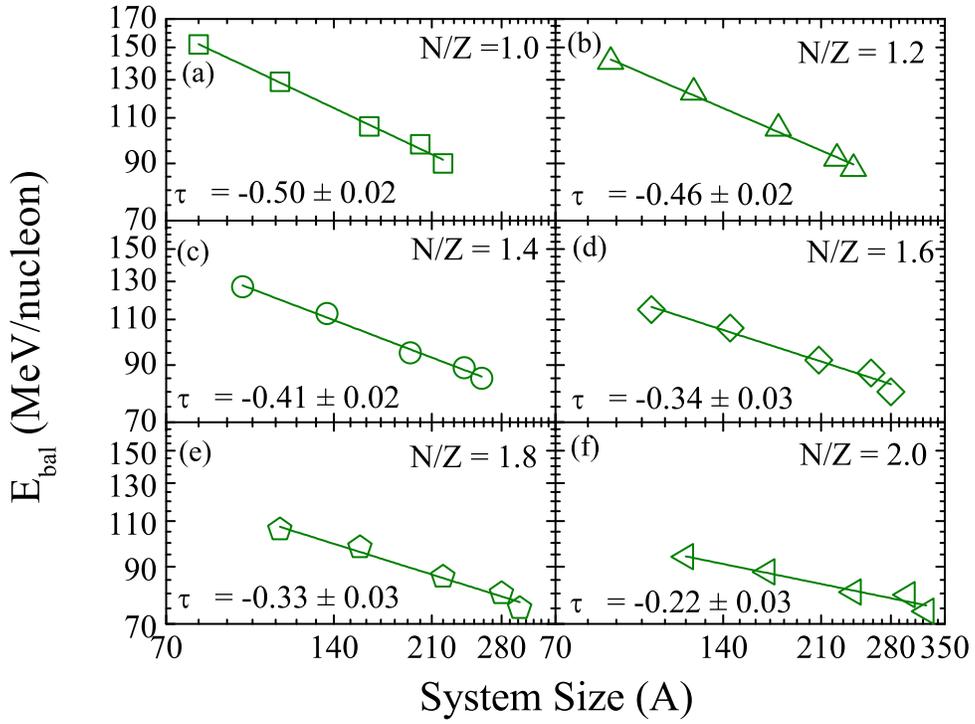}
 \vskip -0cm \caption{System size dependence of E$_{bal}$ for various N/Z ratios at b/b$_{max}$ = 0.4-0.6. Lines are of power law nature ($\propto$ A$^{\tau}$).}\label{fig4}
\end{figure}

\begin{figure}[!t]
\centering
 \vskip 1cm
\includegraphics[angle=0,width=14cm]{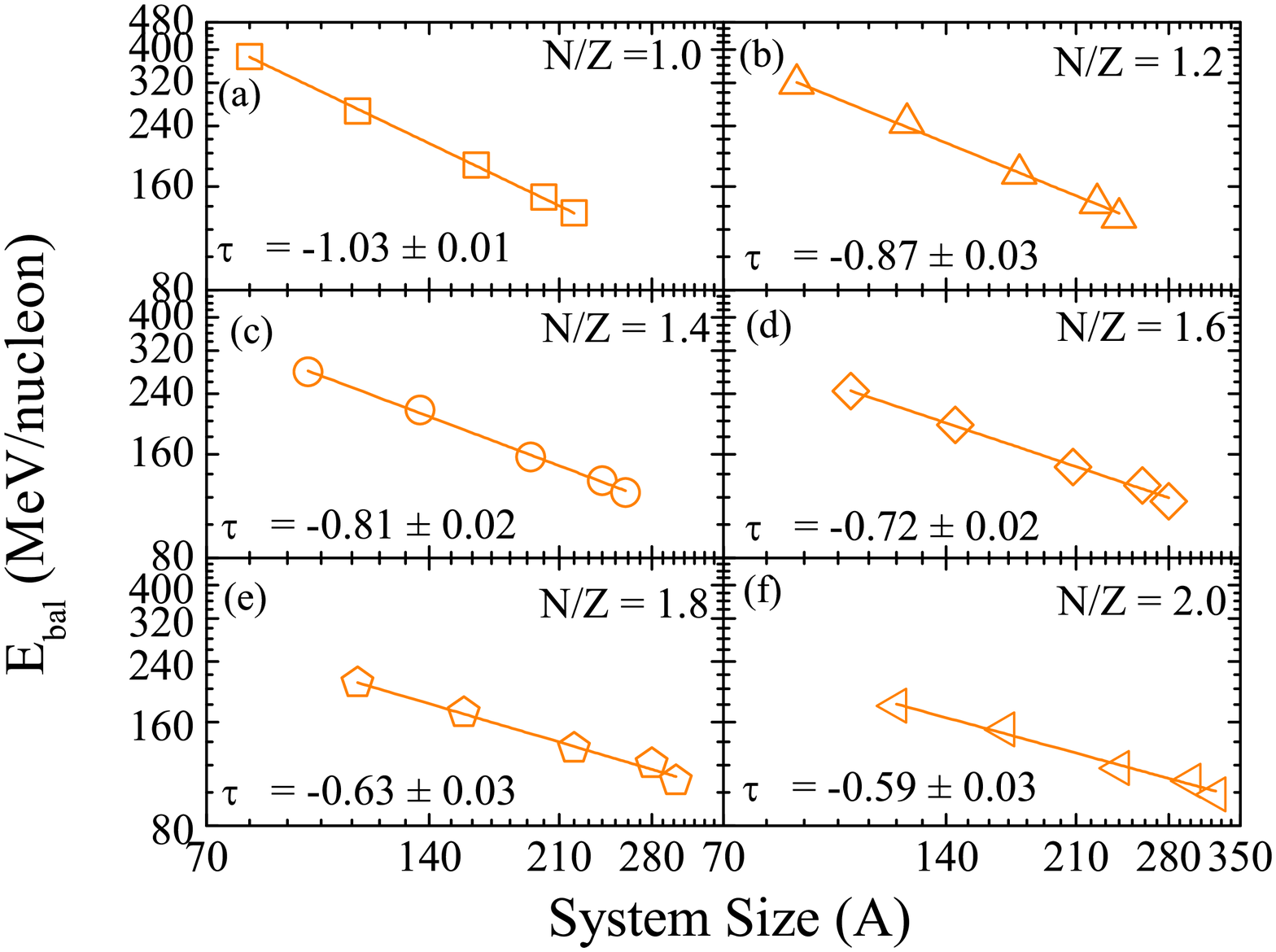}
 \vskip -0cm \caption{ Same as fig. 4 but for b/b$_{max}$ = 0.6-0.8.}\label{fig5}
\end{figure}

In fig. 1 we display the N/Z dependence of E$_{bal}$ for
b/b$_{max}$ = 0.2-0.4 (top panel), 0.4-0.6 (middle panel) and 0.6
-0.8 (bottom panel). From figure, we find that at all the
colliding geometries E$_{bal}$ follows a linear behaviour with
N/Z. The slopes are 33, 25, 21, 18, and 15 (at b/b$_{max}$ =
0.2-0.4), 58, 41, 27, 19, and 18 (at b/b$_{max}$ = 0.4-0.6) and
187, 115, 67, 39, and 36 (at b/b$_{max}$ = 0.6-0.8)
for the series of Ca, Ni, Zr, Xe and Sn, respectively.  From figure, we find that \\
(i) the N/Z dependence of E$_{bal}$ is steeper for the lighter
systems as compared to the heavier systems at all the colliding geometries, \\
(ii) for a particular isotopic series, the N/Z dependence of
E$_{bal}$ is more at peripheral colliding geometry.\\
(iii) and the change in slope is more for lighter systems as
compared to the heavier systems when we move from central to
peripheral colliding geometries. From figure, we see that for Ca
series, slope increases by almost 400\% when we move from central
to peripheral collisions, whereas for Xe series increase in slope
is almost 150\%.
 \par

 \begin{figure}[!t]
\centering
 \vskip 1cm
\includegraphics[angle=0,width=12cm]{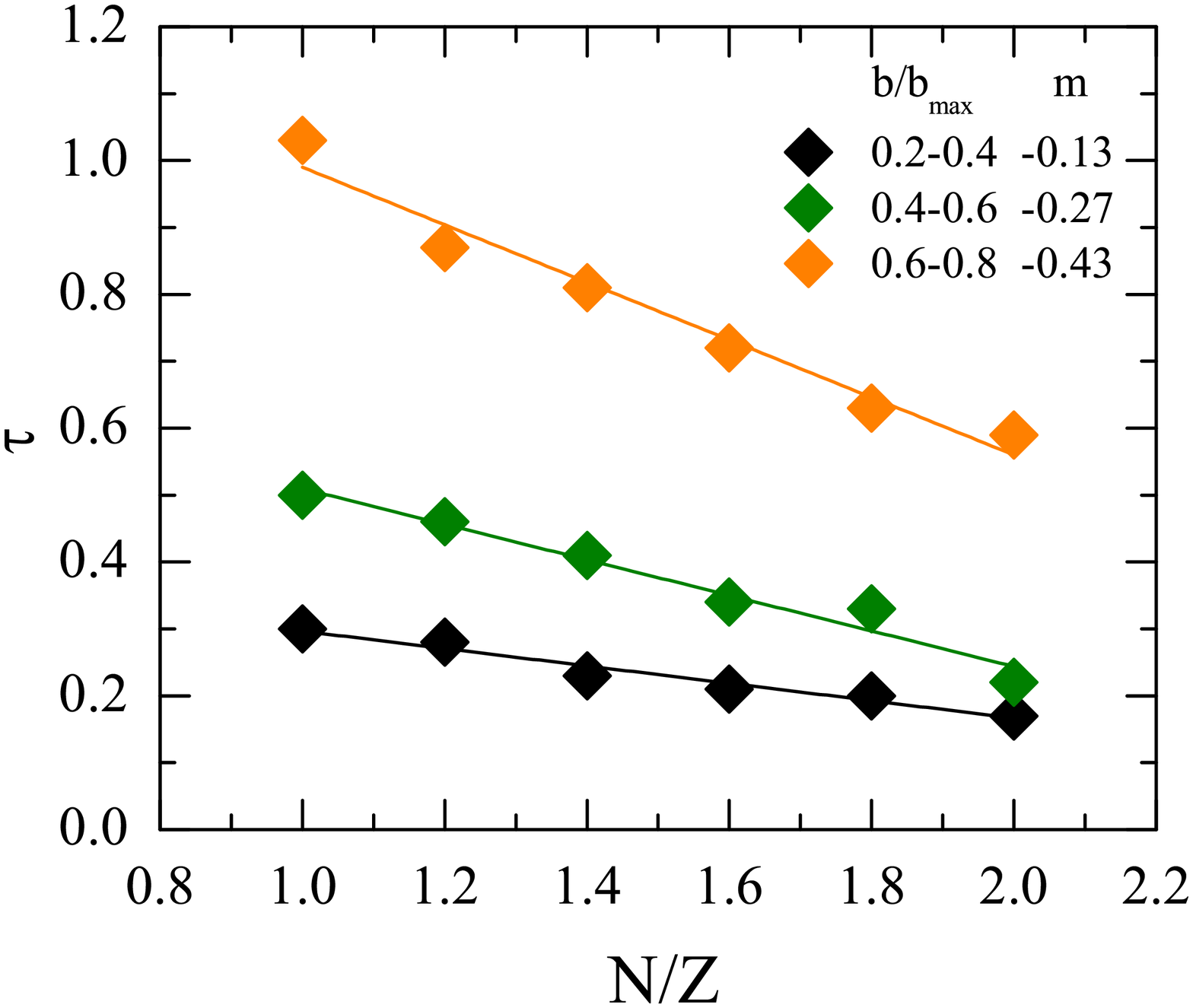}
 \vskip -0cm \caption{ N/Z dependence of $\tau$. Lines are of linear fit.}\label{fig2}
\end{figure}

In Ref. \cite{gaum4} Sood has shown that N/Z dependence of
E$_{bal}$ is sensitive to symmetry energy and is insensitive to
the isospin dependence of nn cross section. The decrease in
E$_{bal}$ with increase in N/Z ratio is due to the enhanced role
of repulsive symmetry energy for higher N/Z ratios. To check the
role of symmetry energy on the N/Z dependence of E$_{bal}$, in
Ref. \cite{gaum5} Gautam and Sood reduces the strength of
potential part of symmetry energy to zero and calculate the
E$_{bal}$. So here also to see the effect of symmetry energy, we
reduce its strength to zero and calculate E$_{bal}$ for two
extreme systems of Ca+Ca and Xe+Xe series. The results are
displayed in Fig. 2 (open symbols). Top, middle and bottom panels
represent the results for  b/b$_{max}$ = 0.2-0.4, 0.4-0.6 and
0.6-0.8, respectively. From figure we see that E$_{bal}$ increases
for both the masses on reducing the strength of symmetry potential
whereas the slope of N/Z dependence of E$_{bal}$ decreases
drastically for both the systems. From figure, we also see that
percentage change ($\Delta$m($\%$)= $\frac{m-m_{symm off}}{m}$) in
slope for Ca (Xe) series is 67 (93) at b/b$_{max}$ = 0.2-0.4
whereas it is 64 (89) at b/b$_{max}$ = 0.4-0.6.
  \par
  In fig. 3, we display the percentage difference of E$_{bal}$ ($\Delta$E$_{bal}$($\%$)= $\frac{E_{bal}^{symm off}
-E_{bal}}{E_{bal}}$*100) between calculations without symmetry
energy and with symmetry energy as a function of N/Z for at
b/b$_{max}$ = 0.2-0.4 (top panel), 0.4-0.6 (middle panel) and
0.6-0.8 (bottom panel). From fig. we
find that\\
(i) the percentage difference increases with N/Z for both the
system masses at all the colliding geometries indicating that the
role of symmetry energy increases with N/Z ratio\\
(ii) and the
increase is more sharp for Ca series as compared to Xe series at
semicentral and semiperipheral collisions whereas at
peripheral collisions, the increase is almost the same for both the masses.\\
(iii) Also, for a particular mass, $\Delta$E$_{bal}$($\%$) rises
with impact parameter and increase with impact parameter is more
for Xe series as compared to the Ca series.
\par
In figs. 4 and fig. 5 we display the system size dependence of
E$_{bal}$ for various N/Z ratios varying from 1.0 to 2.0 for
b/b$_{max}$ = 0.4 - 0.6 and b/b$_{max}$ = 0.6 - 0.8, respectively.
The results are displayed by open squares, triangles, circles,
diamonds, pentagons, and left triangles, for N/Z ratios of 1.0,
1.2, 1.4, 1.6, 1.8, and 2.0, respectively. From fig. 4, we see
that E$_{bal}$ follows a power law behavior ($\propto$ A$^{\tau}$)
with system size. The power law parameter $\tau$ is
-0.50$\pm$0.02, -0.46$\pm$0.02, -0.41$\pm$0.02, -0.34$\pm$0.03,
-0.33$\pm$0.03, and -0.22$\pm$0.03, respectively, for N/Z ratios
of 1.0, 1.2, 1.4, 1.6, 1.8, and 2.0. We find that the power law
parameter goes on decreasing as we are moving towards asymmetric
nuclear matter (higher N/Z). This is due to the fact that for
higher N/Z ratios, the effect of symmetry energy is more in
lighter masses (as in Ref. \cite{gaum5}) and thus decreasing
E$_{bal}$ by larger magnitude in lighter masses which results in
less slope for higher N/Z ratio.
\par
Fig. 5 displays the system size effect of E$_{bal}$ at peripheral
colliding geometry of b/b$_{max}$ = 0.6 - 0.8. The behavior of
E$_{bal}$ with system size is similar, except that now the value
of power law parameter is increased. The value of $\tau$ now reads
as -1.03$\pm$0.01, -0.87$\pm$0.03, -0.81$\pm$0.02, -0.72$\pm$0.02,
-0.63$\pm$0.03, and  -0.59$\pm$0.03,respectively, for N/Z ratios
of 1.0, 1.2, 1.4, 1.6, 1.8,and 2.0. The increase in the value of
slope parameter is due to the fact that E$_{bal}$ for lighter
systems (like Ca+Ca) changes drastically with impact parameter but
the change of E$_{bal}$ with impact parameter in heavier masses is
less \cite{mag00} which leads in the increase of slope parameter
at peripheral colliding geometries.
\par
In Fig. 6 we show N/Z dependence of $\tau$ for the impact
parameter bins of b/b$_{max}$ = 0.2 - 0.4 (black hexagons) ($\tau$
values taken from Ref. \cite{gaum4}), b/b$_{max}$ = 0.4 - 0.6
(green hexagons) and b/b$_{max}$ = 0.6 - 0.8 (orange hexagons). We
see that $\tau$ decreases with increase in N/Z ratio for all the
colliding geometries and follows a linear behavior with N/Z having
slopes 0.13, 0.27, and 0.43 for b/b$_{max}$ = 0.2 - 0.4, 0.4-0.6,
and 0.6 - 0.8, respectively. We also find that change in slope is
more when we move from semi peripheral to peripheral geometry as
compared to that when we move from semi central to semi peripheral
collisions.

\section{Summary}
We have studied the role of colliding geometry on N/Z dependence
of balance energy (E$_{bal}$) for isotopic series throughout the
mass range. We found that dependence of E$_{bal}$ on N/Z ratio is
much stronger for peripheral collisions as compared to the central
collisions. Our study also pointed out that lighter systems showed
more sensitivity to the colliding geometry towards N/Z dependence
of E$_{bal}$. We have also studied the mass dependence of
E$_{bal}$ for the N/Z range from 1.0-2.0 for the whole range of
colliding geometry. We found that the mass dependence of E$_{bal}$
varies with the N/Z ratio.

\par
This work is supported by a grant from Centre of Scientific and
Industrial Research (CSIR), Government of India and Indo-French
center vide project no-4101-A, New Delhi, India.


\begin{thebibliography}{999}


\bibitem{rib1} W. Zhan \emph{et al}., Int. J. Mod. Phys. E \textbf{15}, 1941 (2006); see, e.g. http://www.impcas.ac.cn/zhuye/en/htm/247.htm.
\bibitem{rib2} See, e.g., http://www.gsi.de/fair/index e.html; See, e.g., http://ganiinfo.in2p3.fr.research/developments/spiral2.
\bibitem{rib3} Y. Yano, Nucl. Inst. Methods B \textbf{261}, 1009 (2007).


\bibitem{li96} B. A. Li, Z. Ren, C. M. Ko, and S. J. Yennello, Phys. Rev. Lett. \textbf{76}, 4492
(1996); C. Liewen, Z. Fengshou, and J. Genming, Phys. Rev. C {\bf
58}, 2283 (1998); L. Scalone, M. Colonna and M Di Toro, Phys.
Lett. B {\bf 461}, 9 (1999).

\bibitem{pak97} R. Pak \emph{et al}., Phys. Rev. Lett. \textbf{78}, 1022 (1997);\emph{ ibid}.
\textbf{78}, 1026 (1997).

\bibitem{gaum1} S. Gautam \emph{et al}., J. Phys. G: Nucl. Part. Phys.
\textbf{36}, 085102 (2010).


\bibitem{fang00} D. Q. Fang \emph{et al}., Phys. Rev. C\textbf{ 64}, 044610
(2000); G. J. Kunde \emph{et al}., Phys. Rev. Lett. \textbf{77},
2897 (1996); F. S. Zhang et al., Phys. Rev. C 60, 064604 (1999).

\bibitem{wang09} M. C. Wang \emph{et al}., Phys. Rev. C \textbf{79}, 034606
(2009); M. B. Tsang \emph{et al}., Phys. Rev. Lett. \textbf{92},
062701 (2004); Z. Kohley \emph{et al}., Phys. Rev. C \textbf{83},
044601 (2011).


\bibitem{gaum2} S. Gautam and A. D. Sood, Phys. Rev. C
\textbf{82}, 014604 (2010).

\bibitem{gaum3} S. Gautam, A. D. Sood, R. K. Puri, and J. Aichelin, Phys. Rev.
C, \textbf{83}, 014603 (2011).

\bibitem{gaum4}  A. D. Sood, Phys. Rev.
C, in press (2011).


\bibitem{gaum5} S. Gautam and A. D. Sood, Phys. Rev. C, communicated
(2011).

\bibitem{hart98} C. Hartnack \emph{et al}., Eur. Phys. J. A
\textbf{1}, 151 (1998); C. Hartnack and J. Aichelin, Phys. Rev. C
\textbf{49}, 2801 (1994); S. Kumar \textit{et al}., \textit{ibid}.
\textbf{81}, 014611 (2010); \textit{ibid}. \textbf{81}, 014601
(2010).

\bibitem{aichqmd} J. Aichelin, Phys. Rep. {\bf 202},
233 (1991); E. Lehmann, R. K. Puri, A. Faessler, G. Batko, and S.
W. Huang, Phys. Rev. C \textbf{51}, 2113 (1995); Y. K. Vermani
\textit{et al}., J. Phys. G: Nucl. Part. Phys. \textbf{36}, 105103
(2009); \textit{ibid}. \textbf{37}, 115105 (2010); \textit{ibid}.
Eur Phys Lett \textbf{85}, 62001 (2010); \textit{ibid}. Phys. Rev.
C \textbf{79}, 064613 (2009), Nucl. Phys A \textbf{847}, 243
(2010).

\bibitem{vuu} H. Kruse, B. V. Jacak and H. St$\ddot{o}$cker, Phys. Rev. Lett. {\bf 54},
289 (1985); J. J. Molitoris and H. St$\ddot{o}$cker, Phys. Rev. C
{\bf 32}, 346 (R) (1985).

\bibitem{cug} J. Cugnon, T. Mizutani, and J. Vandermeulen, Nucl. Phys. {\bf
A352}, 505 (1981).

\bibitem{sood1}A. D. Sood and R. K. Puri, Phys. Rev. C \textbf{69}, 054612
(2004).
\bibitem{sood2} A. D. Sood and R. K. Puri, Phys. Lett. \textbf{B594}, 260
(2004).

\bibitem{west93} G. D. Westfall \emph{et al}., Phys. Rev. Lett. {\bf 71}, 1986 (1993).

\bibitem{west98}G. D. Westfall, Nucl. Phys. {\bf A630}, 27c (1998).
\bibitem{cuss}  D. Cussol \emph{et al}., Phys. Rev. C {\bf 65}, 044604 (2002).
\bibitem{ogli89} C. A. Ogilvie \emph{et al}., Phys. Rev. C {\bf 40}, 2592 (1989).

\bibitem{mag00} D. J. Magestro, W. Bauer, and G. D. Westfall, Phys.
Rev. C \textbf{62}, 041603(R) (2000); R. Chugh and R. K. Puri,
Phys. Rev. C \textbf{82}, 014603 ( 2010); D. Klakow, G. Welke, and
W. Bauer, Phys. Rev. C \textbf{48}, 1982 (1993).





\end{thebibliography}
\end{document}